\documentclass[a4paper, 11pt]{article}
\usepackage{amsmath}
\usepackage{graphicx}
\usepackage{latexsym}

\def\b{\begin{eqnarray}}
\def\e{\end{eqnarray}}
\def\n{\noindent}

\begin{document}

\begin{center}

{\LARGE\textbf{Inverse Scattering Transform for the Camassa-Holm
equation
\\}} \vspace {10mm} \vspace{1mm} \noindent

{\large \bf Adrian Constantin$^{a,\dag}$}, {\large \bf  Vladimir S.
Gerdjikov$^{b,\ddag}$} and \\ {\large \bf Rossen I.
Ivanov$^{a,\ast,}$}\footnote{On leave from the Institute for Nuclear
Research and Nuclear Energy, Bulgarian Academy of Sciences, Sofia,
Bulgaria.} \vskip1cm \n \hskip-.3cm
\begin{tabular}{c}
\hskip-1cm $\phantom{R^R} ^{a}${\it School of Mathematics, Trinity
College Dublin,}
\\ {\it Dublin 2, Ireland} \\
$\phantom{R^R}^{b}${\it Institute for Nuclear Research and Nuclear
Energy,}\\ {\it Bulgarian Academy of Sciences,} \\
{\it 72 Tzarigradsko chaussee, 1784 Sofia,
Bulgaria} \\
%{\it Tel:  + 353 - 1 - 608 2898 }\\{\it  Fax:  + 353 - 1- 608 2282} \\
\\{\it $^\dag$e-mail: adrian@maths.tcd.ie}
\\{\it $^\ddag$e-mail: gerjikov@inrne.bas.bg}
\\ {\it $^\ast$e-mail: ivanovr@tcd.ie}
\\
\hskip-.8cm
\end{tabular}
\vskip1cm
\end{center}

%\vskip1cm

\begin{abstract}
\n An Inverse Scattering Method is developed for the Camassa-Holm
equation. As an illustration of our approach the solutions
corresponding to the reflectionless potentials are constructed in
terms of the scattering data. The main difference with respect to
the standard Inverse Scattering Transform lies in the fact that we
have a weighted spectral problem. We therefore have to develop
different asymptotic expansions.

%{\bf PACS:} 02.30.Ik, 05.45.Yv, 45.20.Jj, 02.30.Jr

{\bf MSC:} 35P25, 35Q15, 35Q35, 35Q51, 35Q53

{\bf Key Words:} Hamiltonian systems, Integrable Systems, Lax Pair,
Riemann-Hilbert Problem, Solitons.

\end{abstract}

\section{Introduction}
In this introductory section some well known facts about the
Camassa-Holm (CH) equation and the related spectral problem will be
highlighted.  The CH equation \cite{CH93}
\begin{equation}\label{eq1}
 u_{t}-u_{xxt}+2\omega u_{x}+3uu_{x}-2u_{x}u_{xx}-uu_{xxx}=0,
\end{equation}
where $\omega$ is a real constant, gained popularity as a model
describing the unidirectional propagation of shallow water waves
over a flat bottom \cite{CH93, J02, J03a}  as well as that of
axially symmetric waves in a hyperelastic rod \cite{CS2,Dai98}. It
firstly appeared in \cite{FF81} as an equation with a bi-hamiltonian
structure. CH is a completely integrable equation \cite{CH93,CHH94,
BBS98, CM99, C01, L02, GH03}, describing permanent and breaking
waves \cite{CE98, C00}. Its solitary waves are stable solitons if
$\omega
> 0$ \cite{BBS99, CS00, CS02, J03} or peakons if $\omega = 0$
\cite{CH93,CHH94}. CH arises also as an equation of the geodesic
flow for the $H^{1}$ right-invariant metrics on the Bott-Virasoro
group (if $\omega > 0$) and on the diffeomorphism group (if $\omega
= 0$) \cite{M98, CK03, C00, CKKT04}.

The bi-Hamiltonian form of (\ref{eq1}) is \cite{CH93, FF81}:

\begin{equation}\label{eq2}
 m_{t}=-(\partial-\partial^{3})\frac{\delta H_{2}[m]}{\delta m}=-(2\omega \partial +m\partial+\partial m)\frac{\delta H_{1}[m]}{\delta m}.
\end{equation}

\n where \b\label{eq4a} m = u-u_{xx} \e \n and the Hamiltonians are
\b \label{eq2a} H_{1}[m]&=&\frac{1}{2}\int m u dx
 \\\label{eq2b}
H_{2}[m]&=&\frac{1}{2}\int(u^{3}+uu_{x}^{2}+2\omega u^{2})dx. \e

\n The integration is from $-\infty$ to $\infty$ in the case of
Schwartz class functions, and over one period in the periodic case.

There exists an infinite sequence of conservation laws
(multi-Hamiltonian structure) $H_n[m]$, $n=0,\pm1, \pm2,\ldots$,
such that \cite{FS99,R02,L05,I05}
\begin{equation}\label{eq2aa}
 -(\partial-\partial^{3})\frac{\delta H_{n}[m]}{\delta m}=-(2\omega \partial +m\partial+\partial m)\frac{\delta H_{n-1}[m]}{\delta m}.
\end{equation}

The equation (\ref{eq1}) admits a Lax pair \cite{CH93,C01}

\b \label{eq3} \Psi_{xx}&=&\Big(\frac{1}{4}+\lambda
(m+\omega)\Big)\Psi
 \\\label{eq4}
\Psi_{t}&=&\Big(\frac{1}{2\lambda}-u\Big)\Psi_{x}+\frac{u_{x}}{2}\Psi+\gamma\Psi
\e

\n where $\gamma$ is an arbitrary constant.  We will use this
freedom for a proper normalization of the eigenfunctions.

In our further considerations $m$ will be a Schwartz class function,
$\omega >0$ and $m(x,0)+\omega > 0$. Then $m(x,t)+\omega > 0$ for
all $t$ \cite{C01}. For a discussion of the periodic case we refer
to \cite{CM99} and \cite{C98}. Let $k^{2}=-\frac{1}{4}-\lambda
\omega$, i.e.

\b \label{lambda} \lambda(k)= -\frac{1}{\omega}\Big(
k^{2}+\frac{1}{4}\Big).\e

The spectrum of the problem (\ref{eq3}) under these conditions is
described in \cite{C01}. The continuous spectrum in terms of $k$
corresponds to $k$ -- real. The discrete spectrum (in the upper half
plane) consists of finitely many points $k_{n}=i\kappa _{n}$,
$n=1,\ldots,N$ where $\kappa_{n}$ is real and $0<\kappa_{n}<1/2$.

For all real $k\neq 0$ a basis in the space of solutions of
(\ref{eq3}) can be introduced, fixed by its asymptotic when
$x\rightarrow\infty$  \cite{C01}:

\b \label{eq5} \psi_{1}(x,k)&=&e^{-ikx}+o(1), \qquad
x\rightarrow\infty;
 \\\label{eq6}
\psi_{2}(x,k)&=& e^{ikx}+o(1), \qquad x\rightarrow \infty. \e

\n   Another basis can be introduced, fixed by its asymptotic when
$x\rightarrow -\infty$:

\b \label{eq5a} \varphi_{1}(x,k)&=&e^{-ikx}+o(1), \qquad
x\rightarrow -\infty;
 \\\label{eq6a}
\varphi_{2}(x,k)&=& e^{ikx}+o(1), \qquad x\rightarrow -\infty. \e

\n For all real $k\neq 0$ if $\Psi(x,k)$ is  a solution of
(\ref{eq3}), then $\Psi(x,-k)$ is also a solution, thus

\b \label{eq5aa} \varphi_{1}(x,k)=\varphi_{2}(x, -k), \qquad
\psi_{1}(x,k)=\psi_{2}(x, -k).
 \e

\n Due to the reality of $m$ in (\ref{eq3}) for any $k$ we have

\b \label{eq6aa} \varphi_{1}(x,k)=\bar{\varphi}_{2}(x, \bar{k}),
\qquad \psi_{1}(x,k)=\bar{\psi}_{2}(x, \bar{k})
 \e

The vectors of each of the bases are a linear combination of the
vectors of the other basis:
 \b \label{eq7}
\varphi_{i}(x,k)=\sum_{l=1,2}T_{il}(k)\psi_{l}(x,k) \e

\n where the matrix  $T(k)$ defined above is called the scattering
matrix. For real $k\neq 0$, instead of $\varphi_{1}(x,k)$,
$\varphi_{2}(x,k)$, $\psi_{1}(x,k)$, $\psi_{2}(x,k)$ due to
(\ref{eq6aa}), for simplicity we can write correspondingly
$\varphi(x,k)$, $\bar{\varphi}(x,k)$, $\psi(x,k)$,
$\bar{\psi}(x,k)$. Thus $T(k)$ has the form

\b \label{T} T(k) = \left( \begin{array}{cc}
   a(k)&  b(k)  \\
  \bar{b}(k) &  \bar{a}(k) \\
\end{array}  \right) \,
\e

\n and clearly

\b \label{eq8} \varphi(x,k)=a(k)\psi(x,k)+b(k)\bar{\psi}(x,k). \e

\n The Wronskian $W(f_{1},f_{2})\equiv
f_{1}\partial_{x}f_{2}-f_{2}\partial_{x}f_{1}$ of any pair of
solutions of (\ref{eq3}) does not depend on $x$. Therefore

\b \label{eq9} W(\varphi(x,k), \bar{\varphi}(x,k))= W(\psi(x,k),
\bar{\psi}(x,k))=2ik \e

\n From (\ref{eq8}) and (\ref{eq9}) it follows that

\b \label{eq10} |a(k)|^{2}-|b(k)|^{2}=1, \e

\n i.e. $\det (T(k))=1$.

In analogy with the spectral problem for the KdV equation
\cite{ZMNP}, one can see that the quantities
$\mathcal{T}(k)=a^{-1}(k)$ and $\mathcal{R}(k)=b(k)/a(k)$ represent
themselves the transmission and reflection coefficients respectively
\cite{C01,CI06}. From (\ref{eq10}) it follows that the scattering is
unitary, i.e.

\b \label{eq13} |\mathcal{T}(k)|^{2}+|\mathcal{R}(k)|^{2}=1. \e

\n The entire information about $T(k)$  (\ref{T}) is provided by
$\mathcal{R}(k)$ for $k>0$ only \cite{CI06}. It is sufficient to
know $\mathcal{R}(k)$ only on the half line $k>0$, since from
(\ref{eq5aa}) and (\ref{eq8}), $\bar{a}(k)=a(-k)$,
$\bar{b}(k)=b(-k)$ and thus $\mathcal{R}(-k)=\bar{\mathcal{R}}(k)$.

At the points of the discrete spectrum, $a(k)$ has simple zeroes
\cite{C01}, therefore  $\varphi$ and $\bar{\psi}$ are linearly
dependent (\ref{eq8}):

\begin{equation} \label{eq200} \varphi(x,i\kappa_n)=b_n\bar{\psi}(x,-i\kappa_n).
\end{equation}

\n  In other words, the discrete spectrum is simple, there is only
one (real) eigenfunction $\varphi^{(n)}(x)$, corresponding to each
eigenvalue $i\kappa_n$, and we can take this eigenfunction to be

\b \label{eq201}\varphi^{(n)}(x)\equiv \varphi(x,i\kappa_n)\e

\n The asymptotic of $\varphi^{(n)}$, according to (\ref{eq5a}),
(\ref{eq6}), (\ref{eq200}) is

\b \label{eq203} \varphi^{(n)}(x)&=&e^{\kappa_n x}+o(e^{\kappa_n
x}), \qquad x\rightarrow -\infty;
 \\\label{eq204}
\varphi^{(n)}(x)&=& b_n e^{-\kappa_n x}+o(e^{-\kappa_n x}), \qquad
x\rightarrow \infty. \e

\n The sign of $b_n$ obviously depends on the number of the zeroes
of $\varphi^{(n)}$. Suppose that

\b \label{eqN} 0<\kappa_{1}<\kappa_{2}<\ldots<\kappa_{N}<1/2.\e
Then from the oscillation theorem for the Sturm-Liouville problem
\cite{B}, $\varphi^{(n)}$ has exactly $n-1$ zeroes. Therefore

\b \label{eq205} b_n= (-1)^{n-1}|b_n|.\e

The set

\b \label{eq206} \mathcal{S}\equiv\{ \mathcal{R}(k)\quad (k>0),\quad
\kappa_n,\quad |b_n|,\quad n=1,\ldots N\} \e

\n is called scattering data. The Hamiltonians for the CH equation
in terms of the scattering data are presented in \cite{CI06}.

The time evolution of the scattering data can be easily obtained as
follows. From (\ref{eq8}) with $x\rightarrow\infty$ one has

\b \label{eq14} \varphi(x,k)=a(k)e^{-ikx}+b(k)e^{ikx}+o(1). \e The
substitution of $\varphi(x,k)$ into (\ref{eq4}) with
$x\rightarrow\infty$ gives

\b \label{eq15} \varphi_{t}=\frac{1}{2\lambda}\varphi_{x}+\gamma
\varphi \e

\n From (\ref{eq14}), (\ref{eq15}) with the choice
$\gamma=ik/2\lambda$ for the eigenfunction $\varphi(x,k)$ we obtain

\b \label{eq16} \dot{a}(k,t)&=&0,
 \\\label{eq17}
\dot{b}(k,t)&=& \frac{i k}{\lambda }b(k,t), \e

\n where the dot stands for derivative with respect to $t$. Thus
 \b \label{eq18} a(k,t)=a(k,0), \qquad b(k,t)=b(k,0)e^{\frac{i
k}{\lambda }t}; \e

\b \label{eq19} \mathcal{T}(k,t)=\mathcal{T}(k,0), \qquad
\mathcal{R}(k,t)=\mathcal{R}(k,0)e^{\frac{i k}{\lambda }t}. \e

In other words, $a(k)$ is independent on $t$ and will serve as a
generating function of the conservation laws.

The time evolution of the data on the discrete spectrum is found as
follows. $i\kappa_n$ are zeroes of $a(k)$, which does not depend on
$t$, and therefore $\dot {\kappa}_n=0$. From (\ref{eq4}) with
$\gamma=ik/2\lambda$; $k=i\kappa_n$ and (\ref{eq204}) one can obtain
 \b \label{eq207} \dot{b}_n=\frac{4\omega
\kappa_n}{1-4\kappa_n^2}b_n. \e

The Poisson brackets for the scattering data of the Camassa-Holm
equation are computed in \cite{CI06} where also the action-angle
variables are expressed in terms of the scattering data.

In Section 2 we compute the asymptotics for large $k$ of the
scattering data and the eigenfunctions, which we use in Section 3 to
develop the Inverse Scattering Transform for the CH equation.  A
number of recent papers \cite{C01,J03,Li04,Li05} used a Liouville
transformation to reduce the weighted spectral problem (\ref{eq3})
to a standard problem. Our approach is more direct, and provides, we
believe, more transparent formulas. The special case of
reflectionless potentials ($\mathcal{R}(k)=0$ for all $k$) which
corresponds to the important class of solutions, namely the
multi-soliton solutions is separately studied in Section 4. A
formula for the $N$-soliton solution is obtained.

\section{Analytic solutions and Riemann-Hilbert Problem}

For the application of the Inverse Scattering Method it will be
necessary the asymptotics for large $k$ of $a(k)$ and the Jost
solutions to be found. Firstly we compute the asymptotic of $a(k)$.

 The solution of (\ref{eq3}) can be represented in the form
\begin{equation}\label{eqi1}
 \varphi(x,k)=\exp \Big( -ikx + \int _{-\infty}^{x}\chi(y,k)dy
 \Big).
\end{equation}

\n For $\mathrm{Im}\phantom{*} k>0$ and $x\rightarrow \infty$,
$\varphi(x,k)e^{ ikx}=a(k)$, i.e.

\begin{equation}\label{eqi2}
 \ln a(k)= \int _{-\infty}^{\infty}\chi(x,k)dx, \qquad \mathrm{Im}\phantom{*} k>0.
\end{equation}

\n Since $a(k)$ does not depend on $t$, the expressions $\int
_{-\infty}^{\infty}\chi(x,k)dx$ represent integrals of motion for
all $k$. The equation for $\chi(x,k)$ follows from (\ref{eq3}) and
(\ref{eqi1})

\begin{equation}\label{eqi3}
 \chi_x (x,k)+\chi^2-2ik\chi=-\frac{1}{\omega}\Big(k^2+\frac{1}{4}\Big)m(x)
\end{equation}

\n and admits a solution with the asymptotic expansion
\begin{equation}\label{eqi4}
 \chi(x,k)= p_1 k+p_0+\sum_{n=1}^{\infty}\frac{p_{-n}}{k^n}.
\end{equation}

\n The substitution of (\ref{eqi4}) into (\ref{eqi3}) gives the
following quadratic equation for $p_1$:

\begin{equation}\label{eqi5}
 p_1 ^{2} -2ip_1+\frac{m}{\omega}=0,
\end{equation}

\n with solutions

\begin{equation}\label{eqi6}
 p_1=i\Big(1\pm \sqrt{1+\frac{m}{\omega}}\Big)
\end{equation}

\n Since $\int _{-\infty}^{\infty}p_1(x)dx$ is an integral of the CH
equation, presumably finite, we take the minus sign in (\ref{eqi6}).
One can easily see that $p_0$ and all $p_{-2n}$ are total
derivatives \cite{I05} and thus we have the expansion

\begin{equation}\label{eqi7}
 \ln a(k)= -i\alpha k+\sum_{n=1}^{\infty}\frac{I_{-n}}{k^n},
\end{equation}

\n where $\alpha$ is a positive constant (integral of motion):

\begin{equation}\label{eqi8}
 \alpha= \int _{-\infty}^{\infty}\Big(\sqrt{1+\frac{m(x)}{\omega}}-1\Big)dx,
\end{equation}

\n and $I_{-n}=\int _{-\infty}^{\infty}p_{-n}$ are the other
integrals, whose densities, $p_{-n}$ can be obtained reccurently
from (\ref{eqi3}), (\ref{eqi4}) \cite{I05,CI06}.

In terms of the scattering data $\alpha$ can be expressed as
\cite{CI06}

\begin{equation}\label{eqi24}
 \alpha = \sum
 _{n=1}^{N}\ln\Big(\frac{1+2\kappa_n}{1-2\kappa_n}\Big)^2-
 \frac{8}{\pi }\int _{0}^{\infty}\frac{\ln |a(\widetilde{k})|}{4\widetilde{k}^2+1}d\widetilde{k}.
\end{equation}

The asymptotic of $a(k)$ for $\mathrm{Im}\phantom{*} k>0$ and
$|k|\rightarrow\infty$ from (\ref{eqi7}) is $a(k)\rightarrow
e^{-i\alpha k}$, or

\begin{equation}\label{eqi9}
 e^{i\alpha k}a(k)\rightarrow 1, \qquad \mathrm{Im}\phantom{*} k>0,
 \qquad |k|\rightarrow\infty.
\end{equation}

\n When $k$ is in the upper half plane the following expression is
valid \cite{CI06}

\begin{equation}\label{eqi21}
 \ln a(k)=-i\alpha k +\sum
 _{n=1}^{N}\ln\frac{k-i\kappa_n}{k+i\kappa_n}+
 \frac{1}{\pi i}\int _{-\infty}^{\infty}\frac{\ln |a(k')|}{k'-k}dk'.
\end{equation}

Let us now consider the asymptotic of the Jost solutions, starting
for example from $\psi(x,k)$, (\ref{eq5}). One can check that the
asymtotic for $|k|\rightarrow \infty$ has the form

\b\label{eq20} \psi(x,k)&=&e^{-ikx+kG(x)} \eta(x,k)\nonumber\\
\eta(x,k)&=&X_0(x)+\frac{X_1(x)}{k}+\frac{X_2(x)}{k^2}+\ldots,\e

\n where, due to (\ref{eq5}), $G(x)\rightarrow 0$ and
$\eta(x,k)\rightarrow1$ for $x\rightarrow\infty$. The substitution
of (\ref{eq20}) into (\ref{eq3}) gives explicitly $G(x)$, $X_0$,
$X_1$, $\ldots$:

\b\label{eq21}
\psi(x,k)=e^{-ikx+ik\int_{\infty}^x(1-\sqrt{\frac{m(y)+\omega}{\omega}})dy}
\Big[\Big(\frac{\omega}{m(x)+\omega}\Big)^{1/4}+\frac{X_1(x)}{k}+\ldots\Big].\e

\n Introducing the function

\b\label{eq22} \xi(x)=\exp
\Big[x+\int_{\infty}^x(\sqrt{\frac{m(y)+\omega}{\omega}}-1)dy\Big],\e

\n which looks like a deformation of the ordinary exponent,
(\ref{eq21}) can be written as

\b\label{eq23} \psi(x,k)=[\xi(x)]^{-ik}
\Big[\Big(\frac{\xi(x)}{\xi'(x)}\Big)^{1/2}+\frac{X_1(x)}{k}+\frac{X_2(x)}{k^2}+\ldots\Big].\e

Furthermore, the function $\underline{\chi}(x,k)\equiv
\psi(x,k)e^{ikx}$ is analytic for $\mathrm{Im}\phantom{*} k<0$,
\cite{C01}. This follows from the representation

\b\label{eq24}
\underline{\chi}(x,k)=1-\frac{\lambda}{k}\int_{x}^{\infty}\frac{e^{2ik(x-x')}-1}{2i}m(x')\underline{\chi}(x',k)dx'.\e

\n Notice that
$\int_{\infty}^x\Big(\sqrt{\frac{m(y)+\omega}{\omega}}-1\Big)dy$ is
bounded for all values of $x$. Indeed,

\b\nonumber \Big|
\int_{\infty}^x\Big(\sqrt{\frac{m(y)+\omega}{\omega}}-1\Big)dy\Big|=\Big|
\int^{\infty}_x\frac{m(y)dy}{\omega\Big(1+\sqrt{\frac{m(y)+\omega}{\omega}}\Big)}\Big|\leq
\int_{-\infty}^{\infty}\frac{|m(y)|}{\omega}dy < \infty \e

\n since $m(x)$ is a Schwartz class function. Therefore the function
\b \label{eq25}\underline{\psi}(x,k)\equiv \psi(x,k)[\xi(x)]^{ik}\e
is also analytic for $\mathrm{Im}\phantom{*} k<0$.

Similarly,

\b \label{eq26} \underline{\varphi}(x,k) &\equiv&
\varphi(x,k)\exp\Big\{
ik\Big[x+\int_{-\infty}^x\Big(\sqrt{\frac{m(y)+\omega}{\omega}}-1\Big)dy
\Big]\Big\}\nonumber\\
&=&\Big(\frac{\xi(x)}{\xi'(x)}\Big)^{1/2}+\frac{\widetilde{X}_1(x)}{k}+\frac{\widetilde{X}_2(x)}{k^2}+\ldots\e

\n is analytic for $\mathrm{Im}\phantom{*} k>0$.

Multiplying (\ref{eq8}) by $\xi(x)/a(k)$ and using (\ref{eqi8}),
(\ref{eq25}) and (\ref{eq26}) we obtain

\b \label{eq27}
\frac{\underline{\varphi}(x,k)}{e^{ik\alpha}a(k)}=\underline{\psi}(x,k)+\mathcal{R}(k)\bar{\underline{\psi}}(x,k)[\xi(x)]^{2ik}.
\e

\n The function $\underline{\varphi}(x,k)/(e^{ik\alpha}a(k))$ is
analytic for $\mathrm{Im}\phantom{*} k>0$, $\underline{\psi}(x,k)$
is analytic for $\mathrm{Im}\phantom{*} k<0$. Thus, (\ref{eq27})
represents an additive Riemann-Hilbert Problem with a jump on the
real line, given by
$\mathcal{R}(k)\bar{\underline{\psi}}(x,k)[\xi(x)]^{2ik}$.

\section{Integration of the CH equation by the Inverse Scattering Method}

Let us take an arbitrary $k$ from the lower half plane
($\mathrm{Im}\phantom{*} k<0$). Then using the Residue Theorem,
(\ref{eqi8}) and (\ref{eq200}) we can compute the integral \b
\label{eq30} \frac{1}{2\pi
i}\oint_{C_+}\frac{\underline{\varphi}(x,k')}{e^{ik'\alpha}a(k')}\frac{d
k'}{k'-k}&=& \sum_{n=1} ^N
\frac{\underline{\varphi}(x,i\kappa_n)}{(i\kappa_n-k)e^{-\kappa_n\alpha}a'(i\kappa_n)}\nonumber\\
&=&\sum_{n=1} ^N
\frac{b_n[\xi(x)]^{-2\kappa_n}\underline{\psi}(x,-i\kappa_n)}{a'(i\kappa_n)(i\kappa_n-k)},\e

\n where $C_+$ is the closed contour in the upper half plane
(Fig.1).
\begin{figure}{}\label{fig:1}
\caption{The contours $\Gamma _\pm $}
%\input{b03-1.mtx}
% GNUPLOT: LaTeX picture with emtex specials
%\setlength{\unitlength}{0.240900pt}
\setlength{\unitlength}{0.160900pt}
\ifx\plotpoint\undefined\newsavebox{\plotpoint}\fi
\sbox{\plotpoint}{\rule[-0.175pt]{0.350pt}{0.350pt}}%
\special{em:linewidth 0.7pt}%
\begin{picture}(1875,1800)(0,0)
\put(1772,1649){\makebox(0,0)[c]{$k$}} \put(1772,1649){\circle{100}}
\put(1424,1585){\makebox(0,0)[l]{$\Gamma_{+}$}}
\put(1424,260){\makebox(0,0)[l]{$\Gamma_{-}$}}
\put(264,923){\vector(1,0){1547}} \put(341,953){\vector(1,0){1006}}
\put(1347,953){\line(1,0){387}} \put(1347,892){\line(1,0){387}}
\put(341,892){\vector(1,0){1006}} \put(1037,158){\vector(0,1){1529}}
\put(1386,1518){\special{em:moveto}}
\put(1373,1525){\special{em:lineto}}
\put(1373,1525){\vector(-2,1){0}} \put(1409,923){\vector(1,0){15}}
\put(1386,327){\special{em:moveto}}
\put(1373,320){\special{em:lineto}}
\put(1373,320){\vector(-2,-1){0}}
\put(1733,953){\special{em:moveto}}
\put(1732,974){\special{em:lineto}}
\put(1730,995){\special{em:lineto}}
\put(1727,1016){\special{em:lineto}}
\put(1724,1037){\special{em:lineto}}
\put(1720,1058){\special{em:lineto}}
\put(1716,1079){\special{em:lineto}}
\put(1710,1099){\special{em:lineto}}
\put(1704,1120){\special{em:lineto}}
\put(1698,1140){\special{em:lineto}}
\put(1691,1160){\special{em:lineto}}
\put(1683,1180){\special{em:lineto}}
\put(1675,1199){\special{em:lineto}}
\put(1666,1218){\special{em:lineto}}
\put(1656,1238){\special{em:lineto}}
\put(1646,1256){\special{em:lineto}}
\put(1636,1275){\special{em:lineto}}
\put(1624,1293){\special{em:lineto}}
\put(1612,1310){\special{em:lineto}}
\put(1600,1328){\special{em:lineto}}
\put(1587,1345){\special{em:lineto}}
\put(1574,1361){\special{em:lineto}}
\put(1560,1377){\special{em:lineto}}
\put(1545,1393){\special{em:lineto}}
\put(1530,1408){\special{em:lineto}}
\put(1515,1423){\special{em:lineto}}
\put(1499,1437){\special{em:lineto}}
\put(1483,1451){\special{em:lineto}}
\put(1466,1465){\special{em:lineto}}
\put(1449,1477){\special{em:lineto}}
\put(1432,1490){\special{em:lineto}}
\put(1414,1501){\special{em:lineto}}
\put(1395,1513){\special{em:lineto}}
\put(1377,1523){\special{em:lineto}}
\put(1358,1533){\special{em:lineto}}
\put(1339,1543){\special{em:lineto}}
\put(1319,1552){\special{em:lineto}}
\put(1299,1560){\special{em:lineto}}
\put(1279,1568){\special{em:lineto}}
\put(1259,1575){\special{em:lineto}}
\put(1239,1581){\special{em:lineto}}
\put(1218,1587){\special{em:lineto}}
\put(1197,1592){\special{em:lineto}}
\put(1176,1597){\special{em:lineto}}
\put(1155,1601){\special{em:lineto}}
\put(1134,1604){\special{em:lineto}}
\put(1113,1607){\special{em:lineto}}
\put(1091,1609){\special{em:lineto}}
\put(1070,1610){\special{em:lineto}}
\put(1048,1610){\special{em:lineto}}
\put(1027,1610){\special{em:lineto}}
\put(1005,1610){\special{em:lineto}}
\put(984,1609){\special{em:lineto}}
\put(962,1607){\special{em:lineto}}
\put(941,1604){\special{em:lineto}}
\put(920,1601){\special{em:lineto}}
\put(899,1597){\special{em:lineto}}
\put(878,1592){\special{em:lineto}}
\put(857,1587){\special{em:lineto}}
\put(836,1581){\special{em:lineto}}
\put(816,1575){\special{em:lineto}}
\put(796,1568){\special{em:lineto}}
\put(776,1560){\special{em:lineto}}
\put(756,1552){\special{em:lineto}}
\put(736,1543){\special{em:lineto}}
\put(717,1533){\special{em:lineto}}
\put(698,1523){\special{em:lineto}}
\put(680,1513){\special{em:lineto}}
\put(661,1501){\special{em:lineto}}
\put(643,1490){\special{em:lineto}}
\put(626,1477){\special{em:lineto}}
\put(609,1465){\special{em:lineto}}
\put(592,1451){\special{em:lineto}}
\put(576,1437){\special{em:lineto}}
\put(560,1423){\special{em:lineto}}
\put(545,1408){\special{em:lineto}}
\put(530,1393){\special{em:lineto}}
\put(515,1377){\special{em:lineto}}
\put(501,1361){\special{em:lineto}}
\put(488,1345){\special{em:lineto}}
\put(475,1328){\special{em:lineto}}
\put(463,1310){\special{em:lineto}}
\put(451,1293){\special{em:lineto}}
\put(439,1275){\special{em:lineto}}
\put(429,1256){\special{em:lineto}}
\put(419,1238){\special{em:lineto}}
\put(409,1218){\special{em:lineto}}
\put(400,1199){\special{em:lineto}}
\put(392,1180){\special{em:lineto}}
\put(384,1160){\special{em:lineto}}
\put(377,1140){\special{em:lineto}}
\put(371,1120){\special{em:lineto}}
\put(365,1099){\special{em:lineto}}
\put(359,1079){\special{em:lineto}}
\put(355,1058){\special{em:lineto}}
\put(351,1037){\special{em:lineto}}
\put(348,1016){\special{em:lineto}}
\put(345,995){\special{em:lineto}}
\put(343,974){\special{em:lineto}}
\put(342,953){\special{em:lineto}}
\sbox{\plotpoint}{\rule[-0.350pt]{0.700pt}{0.700pt}}%
\special{em:linewidth 0.7pt}%
\put(342,892){\special{em:moveto}}
\put(343,871){\special{em:lineto}}
\put(345,850){\special{em:lineto}}
\put(348,829){\special{em:lineto}}
\put(351,808){\special{em:lineto}}
\put(355,787){\special{em:lineto}}
\put(359,766){\special{em:lineto}}
\put(365,746){\special{em:lineto}}
\put(371,725){\special{em:lineto}}
\put(377,705){\special{em:lineto}}
\put(384,685){\special{em:lineto}}
\put(392,665){\special{em:lineto}}
\put(400,646){\special{em:lineto}}
\put(409,627){\special{em:lineto}}
\put(419,607){\special{em:lineto}}
\put(429,589){\special{em:lineto}}
\put(439,570){\special{em:lineto}}
\put(451,552){\special{em:lineto}}
\put(463,535){\special{em:lineto}}
\put(475,517){\special{em:lineto}}
\put(488,500){\special{em:lineto}}
\put(501,484){\special{em:lineto}}
\put(515,468){\special{em:lineto}}
\put(530,452){\special{em:lineto}}
\put(545,437){\special{em:lineto}}
\put(560,422){\special{em:lineto}}
\put(576,408){\special{em:lineto}}
\put(592,394){\special{em:lineto}}
\put(609,380){\special{em:lineto}}
\put(626,368){\special{em:lineto}}
\put(643,355){\special{em:lineto}}
\put(661,344){\special{em:lineto}}
\put(680,332){\special{em:lineto}}
\put(698,322){\special{em:lineto}}
\put(717,312){\special{em:lineto}}
\put(736,302){\special{em:lineto}}
\put(756,293){\special{em:lineto}}
\put(776,285){\special{em:lineto}}
\put(796,277){\special{em:lineto}}
\put(816,270){\special{em:lineto}}
\put(836,264){\special{em:lineto}}
\put(857,258){\special{em:lineto}}
\put(878,253){\special{em:lineto}}
\put(899,248){\special{em:lineto}}
\put(920,244){\special{em:lineto}}
\put(941,241){\special{em:lineto}}
\put(962,238){\special{em:lineto}}
\put(984,236){\special{em:lineto}}
\put(1005,235){\special{em:lineto}}
\put(1027,235){\special{em:lineto}}
\put(1048,235){\special{em:lineto}}
\put(1070,235){\special{em:lineto}}
\put(1091,236){\special{em:lineto}}
\put(1113,238){\special{em:lineto}}
\put(1134,241){\special{em:lineto}}
\put(1155,244){\special{em:lineto}}
\put(1176,248){\special{em:lineto}}
\put(1197,253){\special{em:lineto}}
\put(1218,258){\special{em:lineto}}
\put(1239,264){\special{em:lineto}}
\put(1259,270){\special{em:lineto}}
\put(1279,277){\special{em:lineto}}
\put(1299,285){\special{em:lineto}}
\put(1319,293){\special{em:lineto}}
\put(1339,302){\special{em:lineto}}
\put(1358,312){\special{em:lineto}}
\put(1377,322){\special{em:lineto}}
\put(1395,332){\special{em:lineto}}
\put(1414,344){\special{em:lineto}}
\put(1432,355){\special{em:lineto}}
\put(1449,368){\special{em:lineto}}
\put(1466,380){\special{em:lineto}}
\put(1483,394){\special{em:lineto}}
\put(1499,408){\special{em:lineto}}
\put(1515,422){\special{em:lineto}}
\put(1530,437){\special{em:lineto}}
\put(1545,452){\special{em:lineto}}
\put(1560,468){\special{em:lineto}}
\put(1574,484){\special{em:lineto}}
\put(1587,500){\special{em:lineto}}
\put(1600,517){\special{em:lineto}}
\put(1612,535){\special{em:lineto}}
\put(1624,552){\special{em:lineto}}
\put(1636,570){\special{em:lineto}}
\put(1646,589){\special{em:lineto}}
\put(1656,607){\special{em:lineto}}
\put(1666,627){\special{em:lineto}}
\put(1675,646){\special{em:lineto}}
\put(1683,665){\special{em:lineto}}
\put(1691,685){\special{em:lineto}}
\put(1698,705){\special{em:lineto}}
\put(1704,725){\special{em:lineto}}
\put(1710,746){\special{em:lineto}}
\put(1716,766){\special{em:lineto}}
\put(1720,787){\special{em:lineto}}
\put(1724,808){\special{em:lineto}}
\put(1727,829){\special{em:lineto}}
\put(1730,850){\special{em:lineto}}
\put(1732,871){\special{em:lineto}}
\put(1733,892){\special{em:lineto}}
\end{picture}
\end{figure}
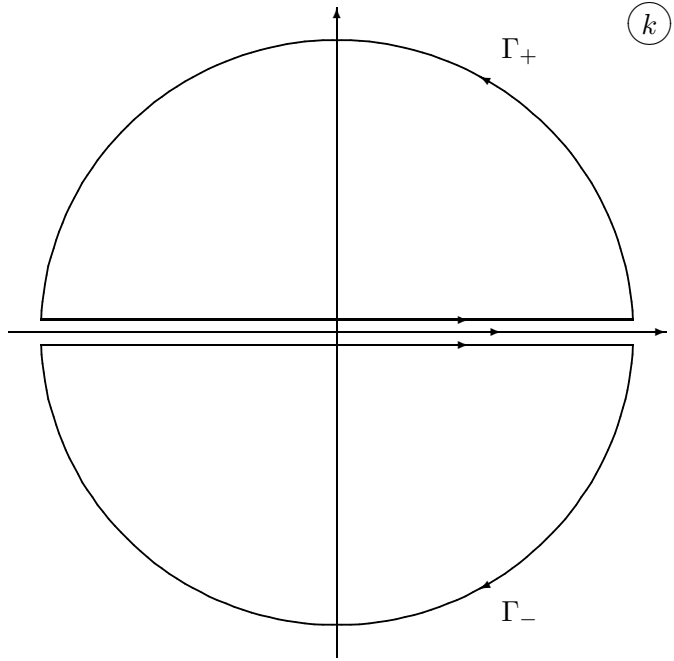
On the other hand, due to (\ref{eq27}) the same integral can be
computed directly as

\b \label{eq31} \frac{1}{2\pi
i}\oint_{C_+}\frac{\underline{\varphi}(x,k')}{e^{ik'\alpha}a(k')}\frac{d
k'}{k'-k}\phantom{***************************}\nonumber \\ =
\frac{1}{2\pi
i}\int_{-\infty}^{\infty}\Big(\underline{\psi}(x,k')+\mathcal{R}(k')\bar{\underline{\psi}}(x,k')[\xi(x)]^{2ik'}\Big)\frac{d
k'}{k'-k}\nonumber \\+\frac{1}{2\pi
i}\int_{\Gamma_+}\frac{\underline{\varphi}(x,k')}{e^{ik'\alpha}a(k')}\frac{d
k'}{k'-k},\e

\n where $\Gamma_+$ is the infinite semicircle in the upper half
plane (Fig.1). Using the expansion (\ref{eq26}) and (\ref{eqi9}), it
is straightforward to compute that the integral over $\Gamma_+$ is
simply $(1/2)(\xi(x)/\xi'(x))^{1/2}$.

Similarly,

\b \label{eq32} -\underline{\psi}(x,k)&=&\frac{1}{2\pi
i}\oint_{C_-}\underline{\psi}(x,k')\frac{d k'}{k'-k}\nonumber \\ &=&
\frac{1}{2\pi i}\int_{-\infty}^{\infty}\underline{\psi}(x,k')\frac{d
k'}{k'-k}+\frac{1}{2\pi
i}\int_{\Gamma_-}\underline{\psi}(x,k')\frac{d k'}{k'-k},\e where
$C_-$ is the closed contour in the lower half plane, $\Gamma_-$ is
the infinite semicircle in the lower half plane (Fig.1). Due to
(\ref{eq23}), (\ref{eq25}) the integral over $\Gamma_-$ is
$(1/2)(\xi(x)/\xi'(x))^{1/2}$.

Now, from (\ref{eq30}) -- (\ref{eq32}) it follows that for
$\mathrm{Im}\phantom{*} k<0$,

\b \label{eq33}
\underline{\psi}(x,k)=\Big(\frac{\xi(x)}{\xi'(x)}\Big)^{1/2}+\int_{-\infty}^{\infty}\mathcal{R}(k')\bar{\underline{\psi}}(x,k')[\xi(x)]^{2ik'}\frac{d
k'}{k'-k}\nonumber \\ +\sum_{n=1} ^N
\frac{b_n[\xi(x)]^{-2\kappa_n}\underline{\psi}(x,-i\kappa_n)}{a'(i\kappa_n)(k-i\kappa_n)}.\e

The expression (\ref{eq33}), taken at $k=-i\kappa_p$, $p=1,\ldots,N$
gives

\b \label{eq34}
\underline{\psi}(x,-i\kappa_p)=\Big(\frac{\xi(x)}{\xi'(x)}\Big)^{1/2}+\int_{-\infty}^{\infty}\mathcal{R}(k')
\bar{\underline{\psi}}(x,k')[\xi(x)]^{2ik'}\frac{d
k'}{k'+i\kappa_p}\nonumber
\\ +i\sum_{n=1} ^N
\frac{b_n[\xi(x)]^{-2\kappa_n}\underline{\psi}(x,-i\kappa_n)}{a'(i\kappa_n)(\kappa_p+\kappa_n)}.\e

The equations (\ref{eq33}) -- (\ref{eq34}) represent a linear
system, from which $\underline{\psi}(x,k)$ (for real $k$) and
$\underline{\psi}(x,-i\kappa_n)$ can be expressed through $\xi$,
which, indeed is yet an unknown function.

Let us now recall that $\lambda(-i/2)=0$. Since $\psi(x,k)$ does not
depend on $m$ for $\lambda=0$ and since $\psi(x,k)$ is defined by
its asymptotics at $-\infty$, it follows that
$\psi(x,-i/2)=e^{-x/2}$. Thus, for $k=-i/2$, (\ref{eq33}) gives

\b \label{eq35}
e^{-x/2}[\xi(x)]^{1/2}=\Big(\frac{\xi(x)}{\xi'(x)}\Big)^{1/2}+\int_{-\infty}^{\infty}\mathcal{R}(k')\bar{\underline{\psi}}(x,k')[\xi(x)]^{2ik'}\frac{d
k'}{k'+i/2}\nonumber \\ +i\sum_{n=1} ^N
\frac{b_n[\xi(x)]^{-2\kappa_n}\underline{\psi}(x,-i\kappa_n)}{a'(i\kappa_n)(\kappa_n+1/2)}.\e

Since $\underline{\psi}(x,k)$ and $\underline{\psi}(x,-i\kappa_n)$
are known from (\ref{eq33}) -- (\ref{eq34}), the equation
(\ref{eq35}) is a first order differential equation for $\xi$, which
can be directly integrated to give $\xi(x)$. In other words,
(\ref{eq33}) -- (\ref{eq35}) represent a system of singular integral
equations for $\underline{\psi}(x,k)$,
$\underline{\psi}(x,-i\kappa_n)$ and $\xi(x)$.

Since the time evolution of the scattering data is known
(\ref{eq207}), the dependence on $t$, i.e. $\xi(x,t)$ is also known,
expressed by the scattering data. Thus the set $\mathcal{S}$,
(\ref{eq206}) uniquely determines the potential: from (\ref{eq207})
one obtains

\b \label{eq36}
m(x,t)=\omega\Big[\Big(\frac{\xi_x(x,t)}{\xi(x,t)}\Big)^{2}-1\Big].\e

\section{Reflectionless potentials}

The inverse scattering is simplified in the important case of the
so-called reflectionless potentials, when the scattering data is
confined to the case $\mathcal{R}(k)=0$ for all real $k$. This class
of potentials corresponds to the $N$- soliton solutions of the CH
equation. In this case $b(k)=0$ and $|a(k)|=1$ (\ref{eq13}) and from
(\ref{eqi21}), (\ref{eqi24}) one can easily find that
$ia'(i\kappa_p)$ is real:

\begin{equation}\label{eq37}
 ia'(i\kappa_p) = \frac{1}{2\kappa_p}e^{\alpha\kappa_p}\prod
 _{n\neq p}\frac{\kappa_p-\kappa_n}{\kappa_p+\kappa_n},
\end{equation}

\n where \begin{equation}\label{eq38}
 \alpha = \sum
 _{n=1}^{N}\ln\Big(\frac{1+2\kappa_n}{1-2\kappa_n}\Big)^2.
\end{equation}

\n Thus, $ia'(i\kappa_p)$ has the same sign as $b_n$, (\ref{eq205})
and therefore

\begin{equation}\label{eq39}
 c_n\equiv \frac{b_n}{ia'(i\kappa_p)}>0.
\end{equation}

\n The time evolution of $c_n$ due to (\ref{eq207}) is

 \b \label{eq40} c_n (t)=c_n(0)\exp\Big[\frac{4\omega
\kappa_n}{1-4\kappa_n^2}t\Big]. \e

The equation (\ref{eq34}) represents a linear system of equations
for the quantities $\underline{\psi}(x,-i\kappa_p)$:

\b \label{eq41} \underline{\psi}(x,-i\kappa_p)+\sum_{n=1} ^N
\frac{c_n[\xi(x)]^{-2\kappa_n}}{\kappa_p+\kappa_n}\underline{\psi}(x,-i\kappa_n)=\Big(\frac{\xi(x)}{\xi'(x)}\Big)^{1/2},
\quad p=1,\ldots,N.\e

\n or

\b \label{eq42}
\underline{\psi}(x,-i\kappa_n)=\Big(\frac{\xi(x)}{\xi'(x)}\Big)^{1/2}\Big[A^{-1}B\Big]_n,
\e

\n where

\b \label{eq43} A_{pn}[\xi,t]\equiv
\delta_{pn}+\frac{c_n(t)\xi^{-2\kappa_n}}{\kappa_p+\kappa_n}, \qquad
B\equiv [\underbrace{1,1,\ldots,1}_N]^{t},\e

\n i.e., finally

\b \label{eq43a}
\underline{\psi}(x,-i\kappa_n,t)=\Big(\frac{\xi(x,t)}{\xi_x(x,t)}\Big)^{1/2}\sum_{p=1}^N
A^{-1}_{np}[\xi,t], \quad n=1,\ldots,N.\e

Now (\ref{eq35}) gives (note that from (\ref{eq22})
$\xi(-\infty,t)=0$)

\b \label{eq44} x=X(\xi,t)\equiv\ln
\int_0^{\xi}\Big(1-\sum_{n,p}\frac{c_n(t)\underline{\xi}^{-2\kappa_n}}{\kappa_n+1/2}A^{-1}_{np}[\underline{\xi},t]\Big)^{-2}d\underline{\xi},
\e

\n the time evolution of $c_n$ is known (\ref{eq40}). This
represents an implicit relation from which $\xi$ can be expressed as
a function of $x$ and $t$. Thus, in this case the scattering data
uniquely determine $\xi=\xi(x,t)$ and therefore the potential
$m(x,t)$ (\ref{eq36}).

Using (\ref{eq44}) and (\ref{eq36}) we obtain the $N$-soliton
solution. Indeed, for fixed coordinates $x_0$, $t_0$, since $x$ is a
monotonically increasing function of $\xi$ there is a unique
$\xi_0>0$ (which is treated as a parameter from now on), such that
$x_0=X(\xi_0,t_0)$. Furthermore, we have

\b \label{eq44a} \xi_x=X^{-1}_{\xi}(\xi,t), \e

\n and from here and (\ref{eq36})

\b \label{eq44b} m(x_0,t_0)\equiv
m(X(\xi_0,t_0),t_0)=\omega\Big[\Big(\xi_0
X_{\xi}(\xi_0,t_0)\Big)^{-2}-1\Big];\e

\b \label{eq44c} u(x_0,t_0)&\equiv&u(X(\xi_0,t_0),t_0)\nonumber
\\ &=&\frac{1}{2}\int_{0}^{\infty}e^{-|X(\xi_0,t_0)-X(\underline{\xi},t_0)|}
m(X(\underline{\xi},t_0),t_0)X_{\xi}(\underline{\xi},t_0)d\underline{\xi}\nonumber
\\ &=&\frac{\omega}{2}\int_{0}^{\infty}e^{-|x_0-X(\underline{\xi},t_0)|}
\underline{\xi}^{-2}
X_{\xi}^{-1}(\underline{\xi},t_0)d\underline{\xi}-\omega. \nonumber
\e

Finally, the $N$-soliton solution is

\b \label{eq44d}
u(x,t)=\frac{\omega}{2}\int_{0}^{\infty}e^{-|x-X(\xi,t)|} \xi^{-2}
X_{\xi}^{-1}(\xi,t)d\xi-\omega.\e

\n Note that $X(\xi,t)$ is an explicitly defined function
(\ref{eq44}) in terms of the scattering data. Thus, the solution
(\ref{eq44d}) does not depend on any additional parameter.

For example, for the one-soliton solution the function $X(\xi,t)$
is:

\b \label{eq45} X(\xi,t)=\ln
\int_0^{\xi}\Big[\frac{1+\frac{1}{2\kappa_1}c_1(t)\underline{\xi}^{-2\kappa_1}}
{1+(\frac{1}{2\kappa_1}-\frac{1}{\kappa_1+1/2})c_1(t)\underline{\xi}^{-2\kappa_1}}\Big]^2d\underline{\xi}.
\e

\n Note that since $\frac{1}{2\kappa_1}-\frac{1}{\kappa_1+1/2}>0$
[cf. (\ref{eqN})] and $c_1(t)>0$ [cf. (\ref{eq39})], both the
nominator and the denominator in (\ref{eq45}) are always positive
and singularities do not appear.

\section{Conclusions} \label{sec:1}
In this paper the Inverse Scattering Method for the CH equation is
outlined in the case when the solutions are confined to be functions
in the Schwartz class, $\omega>0$ and $m(x,0)+\omega>0$. The
$N$-soliton solution is explicitly constructed. The inverse
scattering based on a Liouville transform, which leads to a spectral
problem of the standard form $-\Psi_{yy}+Q(y)\Psi=\mu\Psi$ (the
Schr{\"o}dinger equation) is developed in series of works
\cite{C01,J03,Li04,Li05}. In this case $Q$ and $m$ are related
through the Ermakov-Pinney ordinary differential equation. The
construction of the soliton solutions based on the bilinear
representation of the CH equation (Hirota's method) is presented in
\cite{PI, PII, PIII}, see also \cite{Ma05} for a parametric
representation of the $N$-soliton solution. The situation when the
condition $m(x,0)+\omega>0$ on the initial data does not hold is
more complicated and requires separate analysis \cite{K05,B04}. If
$m(x,0)+\omega$ changes sign there are infinitely many positive
eigenvalues accumulating at infinity, and singularities can appear
in finite time in the form of wave breaking ($\inf_{x\in
\mathbf{R}}\{u_x\}\rightarrow - \infty$, while $u$ stays uniformly
bounded), cf.  \cite{C01, CE98}. The inverse scattering for
multi-peakon solutions (for their existence we must have $\omega=0$)
is reported in \cite{BBS98,BBS99}. For the periodic solutions see
\cite{C98,CM99,GH03} (in this setting a scaling transform shows that
there are no qualitative differences between the cases $\omega=0$
and $\omega\neq 0$). The traveling-wave solutions of the CH equation
($\omega=0$) are classified in \cite{PV05}. The Darboux transform
for the CH equation is obtained in \cite{S98}. The construction of
multi-soliton and multi-positon solutions using the
Darboux/B{\"a}cklund transform is presented in \cite{H99,PLA}.

\section*{Acknowledgements}

A.C. acknowledges funding from the Science Foundation Ireland, Grant
04/BR6/M0042. V.S.G. acknowledges funding from the Bulgarian
National Science Foundation, Grant 1410, R.I.I. acknowledges funding
from the Irish Research Council for Science, Engineering and
Technology. The authors are grateful to both referees for useful
comments and suggestions.

\end{document}